\documentclass[final,3p,times]{elsarticle}

\usepackage{amsthm,amsmath}
\usepackage{graphicx}

\newtheorem{claim}{Claim}[section]
\newtheorem{theorem}[claim]{Theorem}
\newtheorem{proposition}[claim]{Proposition}

\newtheorem{example}[claim]{Example}
\newtheorem{corollary}[claim]{Corollary}

\begin{document}

\begin{frontmatter}

\title{Non-Weyl resonance asymptotics for quantum graphs in a magnetic field}
\author[label1,label2]
{Pavel Exner}
\ead{exner@ujf.cas.cz}
\author[label2,label3]
{Ji\v{r}\'{\i} Lipovsk\'{y}\corref{cor1}}
\ead{lipovsky@ujf.cas.cz}
\address[label1]{Doppler Institute for Mathematical Physics and Applied Mathematics, Czech Technical University,
B\v rehov{\'a} 7, 11519 Prague, Czechia}
\address[label2]{Department of Theoretical Physics, Nuclear Physics Institute, Czech Academy of Sciences, 25068 \v{R}e\v{z} near Prague, Czechia}
\address[label3]{Institute of Theoretical Physics, Faculty of Mathematics and Physics, Charles University, V Hole\v{s}ovi\v{c}k\'ach 2, 18000 Prague, Czechia}
\cortext[cor1]{corresponding author}

\date{\today}
\begin{abstract}
We study asymptotical behaviour of resonances for a quantum graph consisting of a finite internal part and external leads placed into a magnetic field, in particular, the question whether their number follows the Weyl law. We prove that the presence of a magnetic field cannot change a non-Weyl asymptotics into a Weyl one and vice versa. On the other hand, we present examples demonstrating that for some non-Weyl graphs the ``effective size'' of the graph, and therefore the resonance asymptotics, can be affected by the magnetic field.
\end{abstract}

\end{frontmatter}

\section{Introduction}

Quantum graphs became object of intense interest in the last two decades, in part because of numerous possible applications especially in solid-state physics, and also as a tool to study various properties of quantum dynamics. The corresponding bibliography is huge, we limit ourselves to mentioning the proceedings volume \cite{AGA}, in particular the paper \cite{Ku2} there, as a guide to further reading.

An often studied class of graphs are those consisting of a finite internal part to which some number of semiinfinite leads is attached. Such graph typically exhibit a family of resonances; for the purpose of this letter the term \emph{resonance} means both true resonances\footnote{While there are various definitions of true resonances, in particular, as singularities of the scattering matrix, or as poles of analytically continued resolvent, in the present context they were demonstrated to be equivalent \cite{EL} so the term can be used without further specifications.} corresponding to pole singularities in the open lower complex halfplane of energy as well as eigenvalues of the graph Hamiltonian embedded in the continuous spectrum. A question which attracted attention recently concerns the resonance behaviour at high energies. For most graphs of the described class the leading term coefficient in this asymptotics is given by $2 \mathcal{V}/\pi$ where $\mathcal{V}$ denotes the sum of lengths of all internal edges, or ``volume'' of the internal part. This is the usual Weyl law\footnote{The factor two comes from a double-counting, recall that resonance poles appear in pairs coming from the symmetry $k\leftrightarrow-\bar k$.} and we will therefore refer to this case as \emph{Weyl}.

In a surprising observation, however, Davies and Pushnitski \cite{DP} demonstrated that there are situations when this not true and a quantum graph has fewer resonances than the Weyl law would predict, in other words, the relevant ``size'' of the graph entering the leading term of the asymptotics is smaller; we will refer to them as \emph{non-Weyl} in the following. It was shown in \cite{DP} that this happens if the the coupling at the graph vertices is the so-called Kirchhoff one, and at least one vertex is \emph{balanced} joining the same number of internal and external edges. In the subsequent paper \cite{DEL} the claim was extended to graphs with general vertex couplings and a mechanism responsible for ``deleting'' of a part of the internal edges was explained.

The question we address ourselves in the present letter is whether and how the high-energy resonance asymptotics can be changed if the graph is placed into a magnetic field. The exact profile of the field is not important. Since the particle is confined to the graph and behaves locally as a one-dimensional one it is only influenced by the component of the vector potential tangent to the graph edge. Moreover, as we will show in the next section, it is only the fluxes through all closed loops of the graph which are relevant for the resonance behaviour.

\section{Preliminaries}

\subsection{Quantum graphs in magnetic field}

In this section we recall a few basic notions about metric graphs and magnetic Schr\"odinger operators on them. Let $\Gamma$ be a metric graph consisting of set of vertices $\{\mathcal{X}_j\}$ and set of edges $\{\mathcal{E}_j\}$ containing $N$ finite edges and $M$ infinite leads. We equip it with a second order differential operator $H$ acting as $-\mathrm{d}^2/\mathrm{d}x^2$ at the infinite leads and as $-(\mathrm{d}/\mathrm{d}x + i A_j(x))^2$ at the internal edges. Here $A_j$ is the tangent component of the vector potential corresponding to the magnetic field; without loss of generality we may assume that it is zero on external leads because on a edge which is not a part of a loop we can easily remove the vector potential by a gauge transformation.

The domain of the indicated Hamiltonian are functions in $W^{2,2}(\Gamma)$ which satisfy the coupling conditions
 $$
  (U_j - I) \Psi_j + i (U_j+I)(\Psi_j ' + i \mathcal{A}_j\Psi_j) = 0
 $$
at the vertex $\mathcal{X}_j$, where $\Psi_j$ and $\Psi_j'$ are the corresponding vectors of functional values and derivatives, respectively, and $\mathcal{A}_j$ is the diagonal matrix whose entries are limit values of the magnetic potential at the vertex for each internal edge entering the vertex $\mathcal{X}_j$ --- at the beginning of the edge with positive sign, at its end with negative one --- and zero for terms corresponding to the halfline leads.

Writing the coupling condition for each vertex of $\Gamma$ separately may not be practical for some purpose, especially if the internal part consists of a larger number of edges. A useful alternative \cite{Ku2, EL} is to describe the coupling on the whole graph using a single block diagonal $(2N+M)\times (2N+M)$ matrix $U$ by the condition
 \begin{equation}
  (U - I) \Psi + i (U+I)(\Psi'+ i \mathcal{A}\Psi)= 0\,;	 \label{coupl}
 \end{equation}
one may think of a ``flower-like'' graph with all the vertices joined together. The topological structure of the original graph $\Gamma$ is now encoded in the block structure of the matrix $U$ consisting of the blocks $U_j$. In a similar way, the  vectors $\Psi$, $\Psi'$ and the matrix matrix $\mathcal{A}= \mathrm{diag\,}(A_1(0),-A_1(l_1), \dots,A_N(0),-A_N(l_N),0,\dots,0)$ are constructed from $\Psi_j$'s, $\Psi_j'$'s and $\mathcal{A}_j$'s, respectively. The self-adjointness of the Hamiltonian constructed in the described way was proved (with a slightly different notation) in Thm.~2.1 of \cite{KS3}.

Using the local gauge transformation $\psi_j (x) \mapsto \psi_j (x) \mathrm{e}^{-i \chi_j (x)}$ with $\chi_j(x)' = A_j(x)$ one can get rid of the explicit dependence of coupling conditions on the magnetic field and arrive thus at the free Hamiltonian with the coupling conditions given by a transformed unitary matrix,
 \begin{equation}
  (U_A - I) \Psi + i (U_A+I)\Psi' = 0\,,\quad U_A := \mathcal{F} U \mathcal{F}^{-1}	\label{coupl2}
 \end{equation}
with $\mathcal{F}= \mathrm{diag\,}(1,\mathrm{exp\,}(i\Phi_1), \dots,1,\mathrm{exp\,}(i\Phi_N),1,\dots,1)$ containing magnetic fluxes $\Phi_j = \int_0^{l_j} A_j(x) \,\mathrm{d}x$.

Treatment of graphs with external leads can be reduced to analysis of the internal part only. The way to do that is to introduce the effective energy-dependent coupling matrix
 \begin{equation}
   \tilde U(k):= U_1 -(1-k) U_2 [(1-k) U_4 - (k+1) I]^{-1} U_3\,,	\label{effective}
 \end{equation}
on the compact part of the graph obtained from $\Gamma$ by ``chopping off'' the leads \cite{EL, DEL}. In the defining relation \eqref{effective}, the matrices $U_i$ are the blocks of the original coupling matrix $U = \left(\begin{smallmatrix}  U_1& U_2\\U_3& U_4 \end{smallmatrix}\right)$, where the $2N\times 2N$ matrix $U_1$ corresponds to the coupling between the internal edges, the $M \times M$ matrix $U_4$ describes the coupling between the external edges and the rectangular matrices $U_2$, $U_3$ couple the two groups of edges. For a magnetic graph with the coupling (\ref{coupl2}) the matrix $U_A$ can be similarly replaced by the effective coupling matrix $\tilde U_A(k)$ on the compact part of the graph defined in the analogous way.

\subsection{High-energy resonance asymptotics}

The main object of our interest here is the resonance count at high energies, that is, the number of resonances in a circle of radius $R$  in the momentum plane centred at the origin, $k = 0$. According to \cite{DP, DEL} this quantity has the following behaviour
 $$
  N(R) = \frac{2}{\pi} \mathcal{W}R +\mathcal{O}(1)\,,\quad \mathcal{W}\leq \mathcal{V}\,,
 $$
as $R\to\infty$; the effective graph size $\mathcal{W}$  is bound from above by $\mathcal{V}$, the sum of lengths of all the internal edges. If those two quantities coincide, $\mathcal{W} = \mathcal{V}$, the resonances follows the \emph{Weyl asymptotics}, otherwise we have a \emph{non-Weyl} situation. A criterion expressed in terms of the effective coupling matrix \eqref{effective} was derived in \cite{DEL} which allows us to tell when each of the two situations occurs.

\begin{theorem}\label{thm-del}
The graph has a non-Weyl asymptotics iff at least one eigenvalue of $\tilde U(k)$ is $\frac{1+k}{1-k}$ or $\frac{1-k}{1+k}$.
\end{theorem}

\section{Resonance asymptotics of magnetic graphs}

Let us now pass to our main topic. First we observe that under unitary transformations which do not mix the internal and external edges the distinction the non-Weyl asymptotics character is preserved.

\begin{theorem}\label{noswitch}
Let $\Gamma$ be a quantum graph with $N$ internal and $M$ external edges which are coupled by the condition (\ref{coupl2}) with a $(2N+M)\times (2N+M)$ unitary matrix $U$. Let further $\Gamma_V$ be a quantum graph obtained from $\Gamma$ by replacing the coupling matrix $U$ by $V^{-1} U V$ where $V= \left(\begin{smallmatrix}  V_1& 0\\0& V_2  \end{smallmatrix}\right)$ is unitary block-diagonal matrix consisting of a $2N \times 2N$ block $V_1$ and an $M \times M$ block $V_2$. Then $\Gamma_V$ has a non-Weyl resonance asymptotics iff $\Gamma$ does.
\end{theorem}
\begin{proof}
Let the matrix $U$ consist of the blocks $U_1$, $U_2$, $U_3$, $U_4$ in the way described above. Then the coupling matrix of the transformed graph $\Gamma_V$ consists respectively of the blocks $V_1^{-1}U_1 V_1$, $V_1^{-1}U_2 V_2$, $V_2^{-1}U_3 V_1$, $V_2^{-1}U_4 V_2$. Consequently, we have $\tilde U_V (k) =  V_1^{-1} U_1 V_1 -(1-k) V_1^{-1} U_2 V_2 [(1-k) V_2^{-1} U_4 V_2 - (k+1) I]^{-1} V_2^{-1} U_3 V_1 = V_1^{-1} \tilde U(k) V_1$ as the effective energy-dependent coupling matrix for $\Gamma_V$, and it has the same eigenvalues as $\tilde U(k)$. Hence, according to Theorem~\ref{thm-del} the character of the asymptotics does not change.
\end{proof}

Using the fact that the coupling-matrix transformation $U\mapsto U_A$ defined by \eqref{coupl2} belongs to the class covered by Theorem \ref{noswitch} we arrive at the following conclusion.

\begin{corollary}
Let $\Gamma$ be a quantum graph with Weyl resonance asymptotics. Then $\Gamma_A$ has also the Weyl asymptotics for any profile of the magnetic field.
\end{corollary}

In other words, the magnetic field alone cannot switch a graph with non-Weyl asymptotics into one with Weyl asymptotics and {\em vice versa}. On the other hand, as following example shows, the magnetic field \emph{can} change the effective size of a non-Weyl graph.

\begin{figure}
   \begin{center}
     \includegraphics{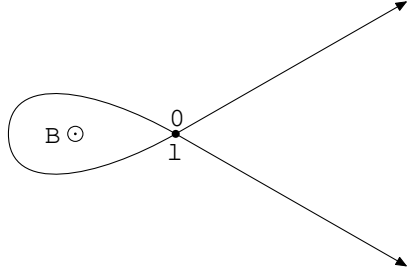}
     \caption{Graph with a balanced vertex connecting two leads with one internal edge}\label{fig1}
   \end{center}
\end{figure}

\begin{example}{\rm
Consider a graph consisting of one vertex connecting a loop of length $l$ and two halflines, with Kirchhoff coupling conditions at the vertex --- cf.~Fig.~\ref{fig1} --- and place it into a magnetic field with constant value of tangent component of the potential. According to \cite{DP} this graph has a non-Weyl asymptotics in the non-magnetic regime. The resonance condition can be easily computed, e.g., by the method of external complex scaling \cite{EL}. Using the Ansatz $f(x) =  \mathrm{e}^{-ikA}(a\mathrm{e^{ikx}}+b\mathrm{e^{-ikx}})$ for the wavefunction on the loop one obtains the set of equations
\begin{eqnarray*}
  ik [a-b-\mathrm{e}^{-iAl}(a \mathrm{e}^{ikl}- b \mathrm{e}^{-ikl})+ 2\mathrm{e}^{-\theta/2}g_\theta (0)]= 0\,,\\
a+b = \mathrm{e}^{-iAl}(a \mathrm{e}^{ikl}+ b \mathrm{e}^{-ikl}) = \mathrm{e}^{-\theta/2} g_\theta (0)\,,
\end{eqnarray*}
where $g_\theta$ denotes the scaled function on the halfline. This leads to the resonance condition
$$
  -2\cos{\Phi} + \mathrm{e}^{-ikl} = 0\,,
$$
where $\Phi = A l$ stands for the magnetic flux through the loop. One can see that for $\Phi  = \pm \pi/2\; (\mathrm{mod}\,\pi)$, that is, for odd multiples of a quarter of the flux quantum $2\pi$, the $l$-independent term disappears.  By the results of \cite{DP,DEL} the effective size of the graph is then zero, i.e. there are finitely many resonances; it is straightforward to see that in the present case there are no resonances at all.

The effective-coupling matrix is in this case equal to
 $$
  \tilde U_\mathcal{A}(k) = \frac{1}{k+1} \begin{pmatrix}  -k & \mathrm{e}^{i\Phi} \\ \mathrm{e}^{-i\Phi} & -k  \end{pmatrix}\,;
 $$
we notice that the resonances are absent when the sum of the off-diagonal elements vanishes.
}\end{example}

The last observation can be generalized to any one-loop quantum graph, with or without a magnetic field.
 \begin{proposition}
The effective size of a graph with a single internal edge is zero iff it is non-Weyl and its effective coupling matrix $\tilde U(k)$ satisfies $\tilde u_{12} + \tilde u_{21} = 0$\,.
 \end{proposition}
 \begin{proof}
According to \cite{DEL} the resonance condition is $F(k)=0$ where
 \begin{multline*}
  F(k) := \mathrm{det\,}\left\{\frac{1}{2}[(\tilde U - I)+k(\tilde U + I)]\begin{pmatrix}0&0\\-i&1\end{pmatrix}\mathrm{e}^{ikl}+
\right.\\ \left.
\frac{1}{2}[(\tilde U - I)-k(\tilde U + I)]\begin{pmatrix}0&0\\i&1\end{pmatrix}\mathrm{e}^{-ikl}
+k(\tilde U+I)\begin{pmatrix}i&0\\0&0\end{pmatrix}+(\tilde U -I)\begin{pmatrix}0&1\\0&0\end{pmatrix}\right\}\,;
 \end{multline*}
for the sake of brevity we dropped here the argument of $\tilde U$. Since the graph is supposed to be non-Weyl, the term with $\mathrm{e}^{ikl}$ in the determinant vanishes, which means that $\tilde U$ has an eigenvalue equal to $\frac{1-k}{1+k}$. To obtain a graph with zero effective size one needs to cancel also the term without the exponentials for which the combination of first two terms in the last displayed equation,
 $$
   -ik \tilde u_{12} (\tilde u_{22}-1) + ik \tilde u_{12} (\tilde u_{22}+1) = 2 ik \tilde u_{12}\,,
 $$
and the last two ones,
 $$
   ik [(\tilde u_{11}+1)\tilde u_{21}+\tilde u_{21}(\tilde u_{11}-1)] = 2ik \tilde u_{21}\,,
 $$
are relevant, hence a necessary and sufficient condition for finiteness of the resonance family is $\tilde u_{12} + \tilde u_{21} = 0$.
 \end{proof}

This leads us directly to the following conclusion.
 \begin{theorem}
For any non-Weyl quantum graph with one internal edge and  $|\tilde u_{12}|  = |\tilde u_{21}|$ there is a magnetic field such that the graph under its influence has at most finite number of resonances.
 \end{theorem}
 \begin{proof}
Under the unitary transformation (\ref{coupl2}) the effective coupling matrix changes to $\left(\begin{smallmatrix}\tilde u_{11}& \mathrm{e}^{i\Phi}\tilde u_{12}\\  \mathrm{e}^{-i\Phi}\tilde u_{21}&\tilde u_{22}\end{smallmatrix}\right)$. Adjusting the phase shift determined by the magnetic flux through the loop one can satisfy the condition of the previous proposition and ensure thus that the effective size of the corresponding magnetic graph is zero.
 \end{proof}

\section*{Acknowledgments}

The research was supported by the Czech Ministry of Education, Youth and Sports within the project LC06002 and by the Charles University in Prague within the project SVV 261301.

\end{document}